\documentclass[12pt,preprint]{aastex}
\usepackage{emulateapj5}



\def\asca       {{\em ASCA}\/}
\def\chandra    {{\em Chandra}\/}

\def\xmm        {XMM-{\em Newton}\/}
\def\rosat      {{\em ROSAT}\/}

\def\mydegree{$^\circ\mskip-5mu$}
\def\myarcmin{$^\prime\mskip-5mu$ }
\def\myarcsec{$^{\prime\prime}\mskip-5mu$}
\def\rv         {{\it r$_{\rm vir}$}\/}
\def\esoa       {ESO 3060170}
\def\esob       {ESO 5520200}

\usepackage{mathptmx}

\begin{document}

\title{\esoa\ --- a massive fossil galaxy group with a heated gas core?}

\author{
M.\ Sun,$^{\!}$\altaffilmark{1}
W.\ Forman,$^{\!}$\altaffilmark{1}
A.\ Vikhlinin,$^{\!}$\altaffilmark{1,2}
A.\ Hornstrup,$^{\!}$\altaffilmark{3}
C.\ Jones,$^{\!}$\altaffilmark{1}
S. S. Murray$^{\!}$\altaffilmark{1}
} 
\smallskip

\affil{\scriptsize 1) Harvard-Smithsonian Center for Astrophysics,
60 Garden St., Cambridge, MA 02138; msun@cfa.harvard.edu}
\affil{\scriptsize 2) Space Research Institute, Russian Academy of Sciences, 84\/32 Profsojuznaya Street, GSP-7, Moscow 117997, Russia.} 
\affil{\scriptsize 3) Danish Space Research Institute,
Juliane Maries Vej 30, Copenhagen O, DK-2100, Denmark.} 

\begin{abstract}

We present a detailed study of the \esoa\ galaxy group combining \chandra,
\xmm\ and optical observations. The system is found to be a fossil galaxy
group. The group X-ray emission is composed of a central dense cool core
(10 kpc in radius) and an isothermal medium beyond the central 10 kpc. The
region between 10 and 50 kpc (the cooling radius) has the same temperature
as the gas from 50 kpc to 400 kpc although the gas cooling time between 10
and 50 kpc (2 - 6 Gyr) is shorter than the Hubble time. Thus, the \esoa\
group does not have a group-sized cooling core. We suggest that the group
cooling core may have been heated by a central AGN outburst in the past
and the small dense cool core is the truncated relic of a previous cooling
core. The \chandra\ observations also reveal a variety of X-ray features
in the central region, including a ``finger'', an edge-like feature and a
small ``tail'', all aligned along a north-south axis, as are the galaxy
light and group galaxy distribution. The proposed AGN outburst may cause
gas ``sloshing'' around the center and produce these asymmetric features.
The observed flat temperature profile to 1/3 \rv\ is not consistent with
the predicted temperature profile in recent numerical simulations. We compare
the entropy profile of the \esoa\ group with those of three other groups and
find a flatter relation than that predicted by simulations involving
only shock heating, S $\propto$ r$^{\sim 0.85}$. This is direct evidence
for the importance of non-gravitational processes in group centers. We
derive the mass profiles within 1/3 \rv\ and find the \esoa\ group is
the most massive fossil group known (1 - 2 $\times$ 10$^{14}$ M$_{\odot}$).
The M/L ratio of the system, $\sim$ 150 at 0.3 \rv, is normal.

\end{abstract}


\section{Introduction}

Historically, optical galaxies have been used to identify and study
a broad range of mass and size scales ranging from galaxy pairs to
superclusters and filaments. Recent X-ray surveys have complemented these
studies and identified a new class of collapsed system, 
Fossil Groups or Over-Luminous Elliptical Galaxies (OLEGs; see Ponman
et al. 1994; Jones, Ponman \& Forbes 2000; Mulchaey \& Zabludoff 1999; Vikhlinin
et al. 1999 for discussions of this unique class) which may represent
the extreme end point of the group merging process and in which the
optical light is dominated by the central galaxy. X-ray studies show
that these optically unremarkable systems are surrounded by extended
dark matter and hot gas halos typical of groups or poor clusters with
spatial extents of up to 1~Mpc (e.g., Vikhlinin et al. 1999). Fossil
groups contribute significantly to the total mass density of the
Universe and are at least as numerous as all poor and rich clusters
combined (of L$_{X} >$ 10$^{43}$ ergs s$^{-1}$) (Vikhlinin et al.
1999; Jones et al. 2003, hereafter J03). 
Fossil groups provide ideal sites to study ``cooling flows'' since
they are generally old and relaxed. Their dark matter structure can be
compared with simulations to test our current understanding of group
evolution. The large M/L ratio of several OLEGS (350 - 600 for h=0.7)
derived by
Vikhlinin et al. (1999) is intriguing since mergers should not alter
a group's mass-to-light ratio. Although there are more than 10 fossil
groups known, few have been studied in detail in X-rays. Recently,
Khosroshahi, Jones \& Ponman (2004) reported the \chandra\ analysis
of a cool fossil group NGC 6482 with a mass at the lower end of the
group scale.

We present in this paper a detailed analysis of a new massive fossil
group \esoa\ using \chandra\ and \xmm\ data. \esoa\ was selected as a
candidate fossil group from a sample of early-type galaxies based on the
\rosat\ All-Sky Survey (RASS; Beuing et al. 1999). The system was classified
as a poor Abell cluster S540 (Abell, Corwin \& Olowin 1989).

Throughout this paper we assume H$_{0}$ = 70 km s$^{-1}$ Mpc$^{-1}$,
$\Omega$$_{\rm M}$=0.3, and $\Omega_{\rm \Lambda}$=0.7. At a redshift
z=0.0358, the luminosity distance to \esoa\ is d$_{\rm L}$ = 158 Mpc, and
1$''$=0.714 kpc. Linear scales, luminosities, densities, entropies,
gas masses, stellar masses, and total masses scale as d, d$^{2}$,
d$^{-1/2}$, d$^{1/3}$, d$^{5/2}$, d$^{2}$, and d, respectively, where
d=d$_{\rm L}$/158 Mpc.

\section{\chandra\ and \xmm\ data analysis}

Throughout the paper, we use the Galactic absorption of 3$\times$10$^{20}$
cm$^{-2}$ for \esoa. The 0.4 - 1.0 keV \xmm\ spectra and 0.6 - 1.0 keV
\chandra\ spectra of \esoa\ can be fitted well by this absorption. The
uncertainties quoted in this paper are 90\% confidence intervals unless
specified otherwise. The solar photospheric abundance table by Anders \& Grevesse
(1989) is used in the spectral fits.

\subsection{\chandra\ observation \& data reduction}

The \chandra\ observations (Table 1) were performed with the Advanced CCD Imaging
Spectrometer (ACIS). The observations were divided into two ACIS-I pointings
(Obs 3188 and 3189),
putting \esoa\ near the center of the I3 and I0 chips respectively.
In each pointing, the optical axis was moved closer to the target along
the detector y-axis. This allows $\sim$ 1$''$ resolution
at the center of \esoa\ and good coverage of the group emission out to
$\sim$ 450 kpc. The data were telemetered in Very Faint mode, which allows
subsequent analysis to reduce soft particle background significantly. \asca\
grades 1, 5 and 7 were excluded, along with the known bad columns, hot pixels,
and CCD node boundaries. We applied the CXC correction
on charge transfer inefficiency (CTI). The slow gain changes in ACIS CCDs
I0-I3 and S2 were also corrected using the program `corr\_tgain' by A. Vikhlinin
\footnote{http://cxc.harvard.edu/contrib/alexey/tgain/tgain.html}.
We investigated the light curves from chip S2 where background is dominant.
A small background flare was found in each pointing and that period of time
was excluded. There is a position offset of $\sim$ 2$''$ between the two
pointings which we corrected using optical identifications of detected X-ray sources.

The CTI-corrected period D blank field background data, 
``acisi\_D\_01236\_bg\_evt\_271103.fits''
\footnote{http://cxc.harvard.edu/contrib/maxim/bg/index.html},
were used. We verified that the \rosat\ soft sky
background flux at the position of \esoa\ matched the average of
the period D blank field background (within 1\%). 
The correction for slow gain changes in ACIS CCDs
I0-I3 and S2 is also applied to the background data. The particle
background levels (measured in PHA channels 2500-3000 ADU for all CCDs)
were 5\% - 6.6\% higher than that of the period D background data.
This is within the range of background variations. The measured 6 - 10
keV fluxes in regions 10$'$ from the group center also show similar
excesses above the level of the period D background. Thus, we increased
the background normalization by 5\% and 6.6\% for Obs 3188
and 3189 respectively to fit the particle background level at the time
of the observations.

Two corrections were made to the ACIS low energy quantum efficiency (QE).
The first corrects for the QE degradation, which increases with time.
The second corrects the QE by an empirical factor of 0.93
below 1.8 keV in the FI CCDs to improve the cross-calibration with the BI
CCDs\footnote{http://asc.harvard.edu/cal/Links/Acis/acis/Cal\_prods/qe/12\_01\_00/}. 
The calibration files used correspond to CALDB 2.26 from the CXC.

\subsection{\xmm\ observation and data analysis}

In the \xmm\ observation (Table 1), the source was put at an offset position (4.5$'$
north of the field center). The medium optical filter was used. Pipeline products
provided by the \xmm\ Science Operations Center (SOC), consisting
of calibrated event files pre-processed with the SAS, were used in this
analysis. We present here the results from the EPIC instruments.
The PN data are largely contaminated by a high background flare so that
2/3 of the exposure is lost. The higher than anticipated EPIC background has
limited our analysis to $\sim$ 0.3 \rv. We used photon events with
patterns 0 to 12 for the MOS data, and 0 to 4 for the PN data.
Since the background data by Lumb et al. (2002) were used, the same flare
rejection criteria as theirs was generally applied. Histograms of high energy
single pixel events ($>$ 10 keV) in time bins of 100 s were made and time
intervals with count rates $>$ 55 (20) events/bin in PN (MOS) are rejected.
We used a higher cut of 55, rather than 45 used in Lumb et al. (2002) for
PN data, to double the useful exposure.

We used EPIC response files: m11\_r7\_im\_all\_2001-11-25.rmf for
MOS1, m21\_r7\_im\_all\_2001-11-25.rmf for MOS2 and epn\_ff20\_sdY0-9.rmf for PN. 
ARF files were generated using SAS ARFGEN.

A double-subtraction method based on the background data produced by
Lumb et al. (2002) is now widely used (e.g., Majerowicz, Neumann \& Reiprich
2002). This method generally involves a $\sim$ 10\% rescaling of
particle background. However, the quiescent background measured in this
observation is significantly higher than those measured by Lumb et al.
(2002): 55\% higher for MOS1 (10 - 12 keV), 36\% higher for MOS2 (10 -
12 keV) and 88\% higher for PN (12 - 14 keV). We also compared the 10 -
14 keV flux with the CLOSED data by Marty et al. (2002). The
out-of-field count rates agree well with the CLOSED data (within 2\%)
but in-field count rates are much higher (30\% - 40\% for MOS data).
All these show that the quiescent background in this observation is
much higher than expected. Thus, a somewhat different method was used
to subtract background.

Since \esoa\ is not a very luminous source, there are source-free
regions in the field that we can use to model the local background.
The surface brightness profile centered on \esoa\ was
first plotted and background regions were identified
by the flattening of the profile at large radii (11.5$'$ - 15.5$'$).
In this (11.5$'$ - 15.5$'$) region, we first subtract Lumb's background.
The residual spectrum is very flat, similar to that of the particle
background. Since this excess particle background is not vignetted, in
the spectral analysis we subtract it in counts/channel space. For each
region, we first subtract Lumb's background. Then, the excess particle
background was subtracted based on the solid angle of each spectral
region. A similar method was used by Churazov et al.
(2003) for the \xmm\ observations of the Perseus Cluster. The background
subtraction for spatial analysis was done in a similar manner. This
method yields results consistent with those derived from the \chandra\ data.

\subsection{X-ray morphology and surface brightness}

The two \chandra\ pointings of \esoa\ were combined and the 0.5 - 4 keV
image (background-subtracted and exposure-corrected) is shown in Fig. 1
as contours superposed on the Digitized Sky Survey (DSS) image. All 
point sources are replaced by surrounding averages. The group emission
is centered on \esoa\ and elongated in the north-south direction with
an ellipticity of $\sim$ 0.3.

The \chandra\ X-ray surface brightness profile of the \esoa\ group (Fig. 2)
shows at least two components separated at $\sim$ 10 kpc. The profile is
fitted by a double-$\beta$ model (Fig. 2): core radius r$_{\rm c}$ =
7.7$^{+2.4}_{-1.5}$ kpc
and $\beta$ = 1.22$^{+0.51}_{-0.26}$ for the inner component; r$_{\rm c}$ =
44.3$^{+1.6}_{-1.4}$ kpc and $\beta$ = 0.514$\pm$0.004 for the outer component
($\chi^{2}$/dof=150.7/23). If the surface brightness profiles of \chandra\
and \xmm\ beyond 10 kpc are jointly fitted by a single $\beta$-model, the
best-fit values are: r$_{\rm c}$ = 47.9$\pm$0.9 kpc and $\beta$ =
0.526$\pm$0.003. The surface brightness profiles steepen at large radii.
Beyond 130 kpc, the surface brightness profiles can be fitted by a power law
with an index corresponding to $\beta$ = 0.558$\pm$0.004.
Based on an offset PSPC pointing of \esoa, we find the surface brightness may
steepen beyond 450 kpc. The surface brightness can be characterized by a power-law
with a slope corresponding to $\beta$=0.58$^{+0.10}_{-0.09}$ between 450 kpc
and 800 kpc.

In view of the obvious ellipticity of the X-ray emission, we also performed
two-dimensional fits, excluding point sources, to the \chandra\ image of the
\esoa\ group using SHERPA. A single elliptical
$\beta$-model fit yields similar results to the one-dimensional fit
(r$_{\rm c}$ = 31.5$\pm$2.0 kpc vs. 34.5$\pm$1.0 kpc; $\beta$ =
0.510$\pm$0.009 vs. 0.491$\pm$0.003). The best-fit ellipticity is
0.280$\pm$0.017, while the position angle is $\sim$ 90\mydegree\ (measured
counterclockwise from west; same hereafter).

\subsection{Global X-ray properties}

The spectra of two ACIS-I pointings, MOS1, MOS2 and PN extracted within 450
kpc of \esoa\ are shown in Fig. 3. The spectra were fitted by a MEKAL model
and the best-fit values are listed in Table 2. The fits are acceptable
and the results from all five different datasets are consistent
with each other. Thus, they were fitted simultaneously.
To determine the abundance of individual elements, we also fit the spectra
with a VMEKAL model. Following Finoguenov, Arnaud \& David (2001), we divide
heavy elements into five groups for fitting: Ne; Mg; Si; S and Ar; Ca, Fe, and
Ni. The best-fit values are: T=2.67$\pm$0.06 keV, Ne=0.67$\pm$0.19,
Mg=0.27$\pm$0.17, Si=0.43$\pm$0.09, S=0.29$\pm$0.13, and Fe=0.48$\pm$0.04
($\chi^{2}$/dof=590.4/505). The derived abundances are typical for hot
galaxy groups (Finoguenov, David \& Ponman 2000). The best-fit VMEKAL models
are also shown in Fig. 3.

We use the relation derived in Evrard, Metzler \& Navarro (1996)
to estimate the virial radius:
\begin{equation}
r_{\rm vir} = 2.78  h_{0.7}^{-1}  (T/10 {\rm keV})^{1/2} (1+z)^{-3/2} {\rm Mpc}
\label{eq:rvir}
\end{equation}
This scaling relation may not apply for cool groups (e.g., Sanderson et al.
2003), but may be used for comparison.
For \esoa, \rv = 1.35 Mpc.

We can estimate the total X-ray luminosity of the \esoa\ group from the
global spectrum. The missing parts (chip edges, chip gaps and point sources)
are accounted for based on the measured surface brightness profile (double
$\beta$-model fit). \chandra\ and \xmm\ results (Table 2) agree within the
current cross-calibration. By averaging these values, we obtain the
rest-frame 0.5 - 2 keV luminosity of 2.6$\times$10$^{43}$ ergs s$^{-1}$
and bolometric luminosity of 6.6$\times$10$^{43}$ ergs s$^{-1}$ within 450
kpc (0.33 \rv).

\subsection{Radial temperature \& abundance profiles}

Although in X-rays, the \esoa\ group is elongated in the north-south direction,
it is useful to derive radially averaged profiles of physical properties
to compare with other groups and numerical simulations. We used a center for all
annuli of $\alpha = 05^{h}$40$^{m}$06$^{s}$.6, $\delta$ =
-40\mydegree~50\myarcmin14\myarcsec (J2000), which is determined from the fits to
X-ray isophotes beyond the central 50 kpc. This position is 4 - 5$''$ south of the
peak of the X-ray distribution and the galaxy nucleus. Radial temperature and
abundance profiles were derived separately for \chandra\ and \xmm\ data. For
ACIS-I data, we required each annulus to contain a total of 1000 - 2000 counts
from the two pointings. For EPIC data, we did spectral fits in 8 broad radial
bins to avoid further correction for the XMM PSF. Point sources are excluded, as well
as chip gaps and boundaries. In the outermost \chandra\ bin (radius 7.2$'$
- 10.5$'$), the group emission still contributes $\sim$ 49\% of flux in the 0.7 -
4 keV band. A low energy cut of 0.7 keV is used to to minimize the effects of
uncertainties in the low energy calibration. Each annulus was fitted by a MEKAL
model, with the temperature and abundance as free parameters. The spectral fits
from the two ACIS-I pointings agree with each other, as do the results from MOS1,
MOS2 and PN. Thus, we performed simultaneous fits to two ACIS-I pointings to
derive \chandra\ profiles, and to MOS and PN data to derive \xmm\ profiles. The
$\chi^{2}_{\nu}$ ranges from 0.8 to 1.2 for \chandra\ spectra (for 21 ---
143 degrees of freedom), and 0.9 to 1.2 for \xmm\ spectra (for 70 --- 401
degrees of freedom), which is acceptable.
The derived temperature and abundance profiles are shown in Fig. 4.
\chandra\ and \xmm\ temperature and abundance profiles are consistent
with each other. Beyond the central cool 10 kpc (the dense cool
core), the temperature profile is flat at $\sim$ 2.7 keV to $\sim$ 0.33 \rv.
We also extracted the spectra of regions outside 450 kpc or 10.5$'$ (mostly
from the S2 chip) which covers only $\sim$ 8\% of the projected area between radii
of 450 - 660 kpc. With the abundance fixed at 0.3 solar (Fig. 4), the best-fit
temperature is 2.5$^{+1.7}_{-0.9}$ keV, which may imply that the temperature
does not significantly decrease at 0.35 - 0.45 \rv.

In view of the non-circular shape of the X-ray isophotes, we also studied
the temperature profiles in four elliptical sectors (north, east, south and
west) beyond the central 10 kpc. The ellipticity of each elliptical bin
is fixed at 0.28 ($\S$2.3) and the position angle is fixed at 90\mydegree.
The azimuthal extents of the north
and south sectors are 71.5\mydegree, while those of the east and west
sectors are 108.5\mydegree. This choice ensures that each sector has the
same area. In each sector, we generated spectra in five annuli. The derived
temperature profiles in all four sectors are flat and agree well with each
other, which again shows that any temperature variation is small beyond
the dense cool core.

\esoa\ is not the only hot group with a flat temperature profile to 0.3 - 0.5
\rv\ measured by \chandra\ and \xmm. Others include A1983 (flat to
0.35 \rv, Pratt \& Arnaud 2003) and
CL1159+5531 (flat to 0.5 \rv, A. Vikhlinin, private communication).
This kind of flat temperature profiles does not agree with the predicted
temperature profile in adiabatic simulations (Loken et al. 2003), which
implies some missing physics in the simulations, e.g., thermal conduction
at 10\% of the Spitzer value (Narayan \& Medvedev 2001; Nath 2003).

Since the temperature profile is flat beyond the central 10 kpc, we only perform
the standard spectral deprojection within the central 20 kpc radius (see Fig. 4).
The result shows that the temperature change across the boundary of the cool
core is abrupt and the temperature gradient inside the cool core is small.

The abundance profile is rather flat beyond the central 40 kpc. At large
radii (200 - 450 kpc), the abundance is $\sim$ 0.35 solar. The regions between
10 and 50 kpc may have the highest abundance. In view of strong lines of Si
and Fe in the spectra, we also measured the radial abundance profiles of Si
and Fe. Between 10 and 50 kpc, the iron abundance (0.88$\pm$0.14) is $\sim$ twice
the abundance in other regions ($<$ 10 kpc and 50 - 450 kpc). Silicon abundances
are poorly constrained but also high between 10 and 50 kpc.

We also investigated the spectra of the central 10 kpc and 10 - 50 kpc regions
to look for signs of multi-phase gas. Both spectra can be well fitted by one
MEKAL model ($\chi^{2}$/dof=70.4/72 and 269.6/251 respectively). We fit the
spectrum with two MEKAL models and a MEKAL model plus a multi-phase cooling-flow
(MKCFLOW) model. Both extra components do not improve the fits. The upper
limits on the mass deposition rate are $\sim$ 2 and 1 M$_{\odot}$ yr$^{-1}$
for the inner 10 kpc and 10 - 50 kpc regions respectively. Thus, we find no
evidence for multi-phase gas within the cooling radius ($\sim$ 50 kpc)
or gas cooler than the ambient medium between 10 and 50 kpc.

\subsection{Electron density, cooling time and entropy profiles}

The electron density profile can be obtained by deprojecting the surface
brightness profile (see Sun et al. 2003 for details, hereafter S03).
The projected flux contribution from regions beyond the outermost bin
is also subtracted based on an offset \rosat\ pointing of \esoa.
From the 0.5 - 4 keV background-subtracted and exposure-corrected surface
brightness profiles, we derived electron density distributions for both
\chandra\ and \xmm\ data as shown in Fig. 5.
The density profile beyond the central 10 kpc can be fitted by a $\beta$-model.
\chandra\ and \xmm\ profiles yield identical fits so that we fit them
simultaneously. The best-fit parameters are:  r$_{\rm c}$ (core radius) =
44.4$\pm$2.6 kpc and $\beta$ = 0.504$\pm$0.011 ($\chi^{2}$/dof=88.7/37). 

With the temperature, abundance and electron density profiles, we derived
the cooling time and entropy (defined as $S = kT/n_{\rm e}^{2/3}$) profiles
(Fig. 6). The cooling time is 2 - 3 $\times$ 10$^{8}$ yr in the very center
and less than a Hubble time ($\sim$ 10$^{10}$ yr) within the central 50 kpc.
The cooling time profile is almost identical to that of the NGC~1550 group
(S03) except for the longer cooling time between 10 and 30 kpc for \esoa\
(Fig. 6). The entropy profile is also similar to the scaled one
\footnote{The scaling factor is (1+z)$^{2}T^{-0.65}$ from Ponman et al. (2003).}
of the NGC~1550 group (S03) within 0.1 \rv\ except for the excess between 10
and 50 kpc in \esoa. Since the entropy can only be changed by cooling or heating,
this implies that the 10 - 50 kpc core region in \esoa\ must have
been heated relative to the cooling core in NGC~1550. At large radii, the
entropy profile rises continuously and shows no sign of flattening.

\section{The structures in the central region}

X-ray and optical images of the central part of the \esoa\ group are shown
in Fig. 7 and physical properties are also shown in Fig. 2, 4, 5 and 6. No
central point-like source is detected and the upper limit on the 0.5 - 10 keV
luminosity is 4 $\times$10$^{40}$ ergs s$^{-1}$ assuming a power law with
an index of 1.7. The
dense cool core within the central 10 kpc is the most significant structure.
One may argue that this component is the hot ISM of \esoa\ based on its location.
However, its X-ray luminosity (7.7$\times$10$^{41}$ ergs s$^{-1}$ in the
0.5 - 2 keV band) places it at the brightest extreme of the galaxy coronae
associated with early-type galaxies and its temperature (1.32$^{+0.08}_{-0.10}$ keV)
is higher than typical values of galaxy coronae (e.g., Brown \& Bregman 1998).
Since this cool core is at the group center and the surrounding medium
has a relatively short cooling time, the properties of this cool core should
be related to those of its surroundings. We discuss its origin in more detail
in $\S$6.

There are X-ray structures found within and beyond the dense cool core
(Fig. 1 and 7) as follows (starting from large scales to small scales):

\begin{description}
\vspace{-0.2cm}
\item[$\bullet$] About 100 kpc from the nucleus, the X-ray emission is more
elongated to the south and shows a steeper
decline to the north. However, at 250 - 300 kpc from the nucleus,
the X-ray isophotes are well centered on \esoa\ with an ellipticity of $\sim$ 0.3.
\vspace{-0.2cm}
\item[$\bullet$] An X-ray ``finger'' (surface brightness extension from
the central peak) extends to the north of the nucleus with a scale of
$\sim$ 30 kpc.
\vspace{-0.2cm}
\item[$\bullet$] While the central dense component shows sharp declines of
surface brightness in all directions, that located $\sim$ 10 kpc north
of the nucleus is the sharpest (edge-like) with a small ``tail'' extending it
to the south.
\end{description}

\vspace{-0.2cm}
The ``finger'' is the most pronounced X-ray feature. Compared to the X-ray
emission at the same radius, the ``finger'' is 20\% - 30\% brighter (significant
at 14 $\sigma$) with $\sim$ 640 net counts (0.4 - 4 keV). The projected temperature
at the position of the ``finger'' is 2.80$^{+0.27}_{-0.30}$ keV and the abundance is
1.03$^{+0.54}_{-0.40}$ solar. Both the gas temperature and abundance are similar
to azimuthally averaged values in the radial range of the ``finger'' (Fig. 4).
If the surroundings of the ``finger'' are used as background, the derived
temperature of the ``finger'' is 2.6$^{+1.1}_{-0.6}$ keV. Thus, the ``finger''
is hotter than the dense cool core and cannot be a cooling wake. Assuming
dimensions of 40$''\times25''\times25''$, its gas mass is $\sim$ 3 $\times 10^{9}$
M$_{\odot}$, comparable to the gas mass of the dense cool core.

The gas density of the ``finger'' is too low and its temperature is too high
to be the accretion wake of \esoa\ (Sakelliou 2000). It is also unlikely a tail
of the stripped material since the the cool core is at the geometric center of
the X-ray isophotes beyond the central 50 kpc (Fig. 7) and the peak of the cool
core is coincident with the nucleus of \esoa. The ``finger'' could arise from
gas motions induced by a previous AGN outburst ($\S$6). During the outburst,
parts of the surrounding gas may be compressed and displaced. With lower
entropy, the displaced gas will eventually fall back. The infall time depends
on the initial offset and should be longer than the free-fall time (several
times 10$^{8}$ yr). Such a central AGN outburst can also explain other central
X-ray structures, the gas motion inside the dense core and the small asymmetric
distribution of X-ray emission at 100 kpc from the nucleus. In the simulations
by Quilis et al. (2001), buoyant bubbles of relativistic particles produced by
a central AGN can induce gas ``sloshing'' (first introduced by Markevitch,
Vikhlinin \& Mazzotta 2001) around the center. The gas ``sloshing'' time depends
on the radius and may produce structures extended in opposite directions on
different scales. This residual gas motion may last for several Gyr.

\section{Optical properties}

\subsection{Group properties and its environment}

We performed optical imaging and spectroscopic observations of the \esoa\
group galaxies with the Danish 1.54m telescope at La Silla, Chile in October
2001, February 2002 and January 2004. Images of the \esoa\ field in Bessel B (900
sec for 13$' \times$ 13$'$) and Bessel R (720 sec for the central 13$' \times 13'$
and 150 sec for a wider 30$' \times 30'$ exposure) bands were obtained. We
have also obtained spectroscopic observations of four galaxies
(\esoa, G2, G3 and G4 in Fig. 1). All these galaxies are group members (Table
3) and have typical early-type galaxy spectra without emission lines. Within
0.5 \rv\ ($\sim$ 16$'$) of \esoa, the second brightest galaxy is 2.6 mag
fainter than \esoa. Thus, \esoa\ is a fossil group based on the definition
of fossil group in J03. In fact, the \esoa\ group is so hot ($\sim$ 2.7 keV)
that it can be considered a poor cluster, making it the most massive
fossil group known. Within \rv,
the second brightest galaxy (ESO 3060160, 19.7$'$ from \esoa) is 1.5 mag fainter
than \esoa\ in the R-band, while all other galaxies are at least 2.4 mag
fainter than \esoa. We estimate that the time for ESO 3060160 to merge into
\esoa\ by dynamical friction is at least 18 Gyr (using equation 7-26
in Binney \& Tremaine, 1987). Thus, any luminous group galaxies
beyond $\sim$ 700 kpc from \esoa\ have not had enough time to merge into \esoa.

The galaxy distribution within \rv\ (32.2$'$) of \esoa\ clearly shows a
north-south elongation, which is in the same direction as that of the
optical light of \esoa. Between \rv\ and 4 \rv\ (the group infall region),
there is a galaxy excess ($\sim$ 60\%) to the north and south. All galaxies
with a similar redshift to \esoa\ (5 in total) in this annulus lie
along the north-south axis. Although detailed optical spectroscopic
observations are required for confirmation, the data support the existence
of a filament extending north-south from 100 kpc to several Mpc.

\subsection{Optical properties of \esoa\ \& evidence of merging}

The optical properties of \esoa\ are important for understanding the group
evolution. We obtained medium resolution (0.83 nm) optical spectra along
along the major and minor axes of \esoa\ with the Danish Faint Object
Spectrograph and Camera on the Danish 1.54m telescope at La Silla, Chile.
The spectra along the major and minor axes are quite similar and we show
the major axis spectrum in Fig. 8. The spectra reveal that \esoa\
is a typical early-type galaxy without recent star-formation activity.
The nuclear spectrum also shows no evidence for emission lines.

The R band galaxy light profile of \esoa\ can be followed to a semi-major
axis of 4.35$'$ or 186 kpc. The measured R-band magnitude of \esoa\ within
that range is 11.72 mag after correction for Galactic extinction and K correction,
which corresponds to an R-band luminosity of 2.63 $\times$ 10$^{11}$ L$_{\odot}$
(M$_{\rm R}$ = -24.4 + 5 log h$_{0.7}$). This is comparable to the
optical luminosity of cDs in rich clusters and somewhat higher than the
average optical luminosity of the brightest galaxies in poor clusters (Fig. 9 of
Thuan \& Romanishin 1981). We fit the R-band optical light with a de
Vaucouleurs r$^{1/4}$ profile and find that the surface brightness declines more rapidly
than the r$^{1/4}$ law. The derived half light radius is 38.5 kpc, but the
best-fit r$^{1/4}$ model overestimates the surface brightness by 30\% -
90\% beyond 60 kpc. The lack of an extended envelope implies \esoa\ is not a cD
galaxy.

We used ELLIPSE in the STSDAS package to measure the ellipticity of \esoa\ in
the R band to a semi-major axis of 2.5$'$ (Fig. 9). The optical ellipticity
is significantly larger than that in X-rays beyond 50 kpc, which is typical
for brightest cluster ellipticals (Porter, Schneider \& Hoessel
1991). The optical position angle is consistent with that in X-rays within
2$'$ (semi-major axis) and changes little ($\sim$ -5\mydegree\ ) beyond
that. The alignment could arise from anisotropic collapse
along the main filament (e.g., West 1994). The orientation of the central galaxy
still reflects the initial density field.

Diffuse and asymmetrical optical light around \esoa\ is revealed after
the model derived by ELLIPSE is subtracted (Fig. 10). The elongated excess
around galaxy G1 may be produced by the tidal force of \esoa.
Similar distortions are also shown in simulations (e.g., Weil, Bland-Hawthorn
\& Malin 1997) as expected in large-mass-ratio accretion. At 130 -
200 kpc south of the nucleus, significant light excess on the level of $\sim$
25.4 mag arcsec$^{-2}$ (R-band) is also detected (Fig. 10). The corresponding
R-band luminosity is $\sim$ 7 $\times$ 10$^{9}$ L$_{\odot}$, about half
that of G1. This region, still along the major axis, is within the tidal
radius of \esoa\ for 10$^{9}$ - 10$^{10}$ M$_{\odot}$ galaxies. Thus,
the southern excess may be the relic of one or several small galaxies
accreting into \esoa. This kind of feature persists for less than 1
Gyr (e.g., Weil et al. 1997), which indicates that \esoa\ is still
accreting small member galaxies along the main filament.

\section{Mass profile \& M/L ratio}

The gas density and temperature profiles can be used to derive the total
gravitational mass profile under the assumption of hydrostatic
equilibrium. The \chandra\ temperature profile beyond the central 10 kpc
and \xmm\ temperature profile beyond the central 40 kpc are fitted
simultaneously with a fifth order polynomial.
To constrain the temperature gradient at $\sim400$ kpc, we included
the temperature measured from a small region between 450 and 660 kpc
(see $\S$2.5) and increased its uncertainty by a
factor of two. The \chandra\ and \xmm\ electron density profiles outside
the dense cool core are modeled by first dividing the profiles with the
best-fit single-$\beta$ model. The residuals are fitted by a sixth-order
polynomial. Combining the temperature and the density profiles and through
a series of Monte Carlo simulations, we derive the total mass profile
outside the central
10~kpc (Fig. 11). The shape of the total mass profile reflects the steepening
of density profile at large radii. Within the central 60 kpc (excluding
the central 15 kpc with large uncertainties), M $\propto$ r$^{2.1\pm0.2}$,
consistent with the prediction of the NFW profile. The gas mass profile can
be derived from the electron density profile. The gas fraction increases
with radius but remains constant at $\sim$ 0.05 between 200 kpc (0.15 \rv)
and 405~kpc (Fig. 11). This gas fraction profile is very similar to that
of A1983 (Pratt \& Arnaud 2003).
 
We measured R-band photometry for galaxies within a 405 kpc
radius. The contribution from background and foreground galaxies is 
estimated from the projected galaxy density to the east and west between
\rv\ and 4 \rv\ radii. Within 405 kpc, about 5\% of galaxies are
background or foreground. The M/L ratio within 405 kpc is $\sim$
150 M$_{\odot}$/L$_{\odot}$, similar to that measured for other systems
(Sanderson \& Ponman 2003). The baryon fraction within 405 kpc is $\sim$
0.08 assuming an R-band stellar mass-to-light ratio of 5.

We can compare the derived total mass distribution to dark matter
halo models (e.g., Navarro, Frenk \& White 1997; Moore et al. 1998).
Although the two parameters in these two profiles, $\delta_{\rm c}$ (central
overdensity) and $r_{\rm s}$ (characteristic radius), are highly degenerate
in fitting the total mass profile over a limited range, the uncertainty
for the concentration parameter is small. The two models fit the data
similarly, but the derived concentration parameter from the Moore profile
($\sim$ 3.3) predicts a reasonable halo mass of $\sim$ 1.8 $\times$ 10$^{14}$
M$_{\odot}$ for the \esoa\ group, while that derived from the NFW profile
($\sim$ 8.7) implies a halo mass $\sim$ 10 times smaller.
Both models predict r$_{200}$ = 0.92 - 1.05 Mpc, which is smaller
than the value predicted by the self-similar relation in EMN96 (1.3 Mpc).

\section{Discussion}

\subsection{The central 50 kpc --- the heated phase of a normal cooling core?}

Although the gas cooling time within the central 50 kpc is smaller than the
Hubble time, this hot group only hosts a small dense cool core (10 kpc in
radius) and lacks a group-sized cooling core which is found in many relaxed
groups (e.g., NGC~1550 - S03; MKW 4 - O'Sullivan et al. 2003). Since fossil
groups are believed to be old and relaxed systems (e.g., Jones et al. 2000),
group-sized cooling cores should develop at their centers, especially when there
is a small dense cool core as the ``seed'' for the inflow of the surrounding
gas (e.g., Brighenti \& Mathews 2002). For the observed parameters of \esoa, a
group-sized cooling core should develop within 2 Gyr in the absence of a
significant heat source.

The puzzle of the missing group-sized cooling core in \esoa\ can be reconciled
with our understanding of fossil groups, if an existing cooling core has been
reheated. Possible heat sources
include: thermal conduction, a central AGN and minor mergers along
the filament. Thermal conduction with $\sim$ 0.2 of the classical Spitzer
value (Narayan \& Medvedev 2001) can just barely balance the cooling between
15 and 50 kpc if the cooling induces a temperature gradient similar to
those in NGC~1550 and MKW 4. The gas cannot be heated to produce the observed,
flat temperature profile. Moreover, the the abrupt temperature change across
the cool core boundary implies a large suppression of thermal conduction.

The energy budget to heat the gas between 10 and 50 kpc to the observed
value is estimated to be $\sim$ 10$^{59}$ ergs. Waves generated by
subgroup infall can provide enough energy to heat the gas in this region
(e.g., Churazov et al. 2003). However, such activity (subgroup infall)
is not apparent in the optical data in which only infall of individual
small galaxies along the filament is present ($\S$4.2). A central
AGN outburst can also provide the required energy. If the central
AGN is active for 10$^{8}$ yr, the required heating rate is $\sim$ 3 $\times
10^{43}$ ergs s$^{-1}$. The PMN surveys (Wright et al. 1994) only give an
upper limit of 48 mJy at 4.85 GHz in the position of \esoa. This region has
been observed by the more sensitive Sydney University Molonglo Sky Survey (SUMSS)
at 843 MHz (Bock, Large \& Sadler 1999) and a 19 mJy source is detected
at the position of \esoa\ (private communication with D. Hunstead). There is
another 19 mJy source detected 2.2$'$ southeast of \esoa\ nucleus, which
may also be associated with \esoa\ in view of the poor angular resolution
of SUMSS (46$'' \times 65''$ at the position of \esoa). This corresponds to
a total radio luminosity (10 MHz - 10 GHz) of 3 - 6 $\times 10^{39}$ ergs
s$^{-1}$ if the spectral index is 0.8. From Fig. 1 of B\^{\i}rzan et al.
(2004), radio sources with this total luminosity can have a
mechanical power on the order of 10$^{43}$ ergs s$^{-1}$. Thus, it is
possible that a central AGN outburst can heat the gas between 10 and 50
kpc. This argument is further strengthened if the central radio source
has faded considerably after the outburst since the lifetime of
the radio synchrotron emission is only 24 (B/10 $\mu$G)$^{-3/2}$
($\nu$/1.4 GHz)$^{-1/2}$ Myr after the injection of relativistic electrons
from the nucleus has stopped.

Therefore, a previous AGN outburst remains as a plausible heat source. In
this scenario, the central 50 kpc, once a typical cooling core, has been
heated by a central AGN outburst in the past. The energy, transported by
narrow jets, was deposited beyond the central 10 kpc so that the central
part of the cooling core survived. The survival of the central cooling
core after an AGN outburst is not a surprise. Bright radio lobes have been
detected around the 1 keV galaxy coronae of NGC~4874 and NGC~3842 (Sun et al.
2004), which strongly implies that most of the mechanical power is released
at moderate distances from the nucleus. After the initial AGN outburst, the
convection mixes the gas between 10 and 50 kpc to reduce the temperature
gradient. In the simulations by Quilis, Bower \& Balogh (2001), about 0.7 Gyr
after the energy injection from bubbles, the displaced gas begins to fall
back toward the core and a cool core may form. Thus, the group-sized cooling
core was truncated at 10 kpc, leaving a dense cool core at the very center.
The magnetic field around the small cool core may have been compressed and
stretched along the boundary so that thermal conduction is largely suppressed
at the boundary of the dense cool core. In this sense, NGC~1550 and \esoa\
possess cooling core at two different phases - a normal cooling core (NGC~1550)
and a heated core after an AGN outburst (\esoa). One group with properties
similar to those of \esoa\ is AWM 4 (O'Sullivan et al. 2004) although there
is no small cool core detected by \xmm.

\subsection{The inner entropy profiles of galaxy groups}

The entropy profile of the \esoa\ group can be compared with those
of other groups. Ponman et al. (2003) studied a sample of 66 virialised
systems and showed that S $\propto$ T$^{\sim 0.65}$, significantly shallower
than the self-similar prediction (S $\propto$ T). We plot the scaled
entropy profiles ((1+z)$^{2}T^{-0.65}S$) of the \esoa\ group and two other
2-3 keV groups in Fig. 12 (ESO 5520200; A1983 - Pratt \& Arnaud 2003), and
one cooler group NGC~1550. The scaled profiles show good agreement from
0.04 to 0.08 \rv, and within 0.008 \rv. Between 0.01 and 0.04 \rv, \esoa\
and \esob\ have entropy excesses relative to the NGC~1550 group. This variety
of inner entropy profiles is related to the evolutionary stages of group
cooling cores. Groups with heated gas cores can have flat inner entropy
profiles (also e.g., AWM 4, O'Sullivan et al. 2004), while those with large
cooling cores do not have flat entropy cores. At large radii, the entropy
profiles of the three 2 - 3 keV groups agree well with each other.

The simulations involving only gravity and shock heating predict
S $\propto$ r$^{1.1}$ (Tozzi \& Norman 2001), which can be tested
by our data. If the flattened entropy profiles from 0.01 - 0.04 \rv\
and the flattened profile of NGC~1550 beyond 0.09 \rv\ are excluded,
our data indicate S $\propto$ r$^{0.85}$. Separate fits to
0.001 - 0.1 \rv\ and 0.1 - 0.3 \rv\ yield S $\propto$
r$^{0.84}$ and S $\propto$ r$^{1.03\pm0.08}$ respectively.
Although the entropy profiles at large radii have a slope close to
that predicted, they certainly flatten within 0.1 \rv. This is an
indication of significant non-gravitational processes at the center.
However, the observed profiles are also inconsistent with recent
simulations including non-gravitational processes (e.g., Borgani et
al. 2004). We also note that the entropy profiles of the hot groups
in our plot are much flatter than that of NGC~1550 beyond the central
cool core. We tried another model composed of a power law and a constant
representing an isentropic core, to fit the entropy
profiles of the hot groups beyond 0.007 \rv. The data can be fitted
well by this ``hybrid'' model. The best-fit power law index is
1.13$\pm$0.07, consistent with the r$^{1.1}$ scaling. Although
this is not exactly the isentropic core shown in the simulations,
it implies that isentropic gas cores can be present in groups with
heated cores.

\section{Conclusion}

We find that the \esoa\ system is a hot (2.7 keV), massive (1 - 2 $\times$
10$^{14}$ M$_{\odot}$) and X-ray luminous (6.6 $\times 10^{43}$ ergs s$^{-1}$)
fossil group. The main conclusions of our study are:

1. The X-ray emission of the group is composed of two components, a central
dense cool core (1.3 keV and 10 kpc in radius) and a hotter ambient medium
($\sim$ 2.7 keV). From 10 kpc to the cooling radius ($\sim$ 50 kpc), the
temperature profile is surprisingly flat, although the gas cooling time is
only 2 - 6 Gyr. Thus, the \esoa\ group does not have a group-sized cooling
core as found in other groups (e.g., NGC~1550, S03; MKW 4, O'Sullivan et al. 2003).
We suggest that the central 50 kpc of the \esoa\ group has been heated
by a central AGN outburst. The cooling core has been disrupted and the
dense cool core within 10 kpc is the relic of the original cooling core.

2. The X-ray emission from the group is elongated north-south,
just as the optical light of \esoa\ and the galaxy distribution of the
group. Some X-ray features are found around the center, including a
``finger'', an edge-like feature and a small tail trailing behind it,
all aligned north-south, which may be caused by gas ``sloshing''
along the north-south filament. The AGN outburst that destroyed the group
cooling core could induce this residual gas motion.

3. \esoa\ is optically luminous (M$_{\rm R}$ = -24.4 + 5
log h$_{0.7}$) and has a normal early-type galaxy optical spectrum. Though no
extended envelope in excess of the de Vaucouleurs profile is found,
asymmetric diffuse optical light is found to 130 - 200 kpc south
of the nucleus, which may be the remains of small galaxies accreted
within the last 1 Gyr.

4. Beyond the cooling radius to 450 kpc (or 1/3 \rv), the temperature
profile is flat (both \chandra\ and \xmm), inconsistent with the
prediction from recent simulations (Loken et al. 2003; Borgani et al.
2004). The abundance profile is rather flat beyond the central 40 kpc.
At large radii (200 - 450 kpc), the abundance is $\sim$ 0.35 solar for
this optically poor group.

5. We compare the scaled entropy profiles (scaled by T$^{-0.65}$) of
three hot groups and one cool group with the simulations.
We find S $\propto$ r$^{\sim 0.85}$, flatter than the prediction
(S $\propto$ r$^{1.1}$) from simulations involving only shock heating.
We suggest that the entropy profiles within the cooling radius are
related to the evolutionary stage of group gas core. Isentropic gas
cores can persist for groups with large heated cores.

6. The gas fraction profile of the \esoa\ group is flat at $\sim$ 0.05
between 200 and 450 kpc. (M/L)$_{\rm R} \sim$ 150 at 405 kpc, which
implies that the mass-to-light ratio of this fossil group is normal.

\acknowledgments

We thank D. Hunstead for providing the SUMSS data of \esoa. We are grateful to E.
Churazov, P. Marty and M. Markevitch for helpful discussions.
We acknowledge support from the Smithsonian Institution and NASA contracts
NAS8-38248 and NAS8-39073.

\vspace{1.0cm}
\begin{table}
\begin{small}
\begin{center}
\caption{X-ray observations}
\begin{tabular}{ccccc}
\hline \hline
Observing Time & Instrument / Telescope & Datamode & Total Exposure (ks) & Effective Exposure (ks)\\
\hline
 Mar. 8, 2002 & ACIS-I / \chandra & VFaint & 14.4 & 14.0 \\
 Mar. 9, 2002 & ACIS-I / \chandra & VFaint & 14.1 & 13.6 \\
 Oct. 11, 2002 & M1, M2, PN / \xmm & PrimeFull & 17.8, 17.8, 14.8 & 15.8, 16.5, 5.2 \\
\hline\hline
\end{tabular}
\end{center}
\end{small}
\end{table}

\begin{table}
\begin{small}
\begin{center}
\caption{The fit to the global spectrum of \esoa\ within 0.33 \rv $^{\rm a}$}
\vspace{0.3cm}
\begin{tabular}{ccccccc}
\hline \hline
 & \chandra\ & \chandra\ & \xmm\ & \xmm\ & \xmm\ & Combined \\
 & Obs 3188 & Obs 3189 & MOS1 & MOS2 & PN & \\\hline

T (keV) & 2.69$\pm$0.10 & 2.57$\pm$0.10 & 2.67$\pm$0.12 & 2.63$\pm$0.11 & 2.57$\pm$0.10 & 2.63$\pm$0.05 \\
Abundance (solar) & 0.47$\pm$0.08 & 0.39$\pm$0.06 & 0.48$\pm$0.08 & 0.44$\pm$0.07 & 0.44$\pm$0.07 & 0.44$\pm$0.03 \\
$\chi^{2}$/dof & 137.0/116 & 121.5/117 & 106.1/84 & 116.4/87 & 116.5/97 & 602.4/509 \\
L$_{\rm X}$ (0.5 - 2 keV)$^{\rm b}$ & 2.47 & 2.55 & 2.41 & 2.65 & 2.69 & 2.55$\pm$0.12 \\
L$_{\rm bol}^{\rm b}$ & 6.48 & 6.66 & 6.20 & 6.79 & 6.87 & 6.60$\pm$0.27 \\
\hline\hline
\end{tabular}
\begin{flushleft}
\leftskip 70pt
$^{\rm a}$ Using MEKAL model and fixing absorption at 3$\times$10$^{20}$ cm$^{-2}$ \\
$^{\rm b}$ Integrated within 450 kpc radius and in unit of 10$^{43}$ ergs s$^{-1}$ \\
\end{flushleft}
\end{center}
\end{small}
\end{table}

\begin{table}
\begin{small}
\begin{center}
\caption{Velocities of the member galaxies}
\begin{tabular}{ccccc}
\hline \hline
Galaxy & RA & decl. & Velocity & R\\
       & (J2000) & (J2000) & (km/s) & (mag) \\ \hline
\esoa\ & 05:40:06.67 & -40:50:11.4 & 10755$\pm$30 & 11.72 \\
G2 & 05:40:12.94 & -40:53:09.5 & 12030$\pm$35 & 14.33 \\
G3 & 05:39:57.30 & -40:53:55.8 & 10823$\pm$36 & 14.78 \\
G4 & 05:40:07.48 & -40:47:09.6 & 9312$\pm$48  & 15.08 \\
ESO 3060160$^{\rm a}$ & 05:39:53.44 & -40:30:45.0 & 11182 & 13.20 \\
\hline\hline
\end{tabular}
\begin{flushleft}
\leftskip 70pt
$^{\rm a}$ The velocity and R magnitude are from NED. \\
\end{flushleft}
\end{center}
\end{small}
\end{table}

\begin{figure*}
  \centerline{\includegraphics[height=1.1\linewidth]{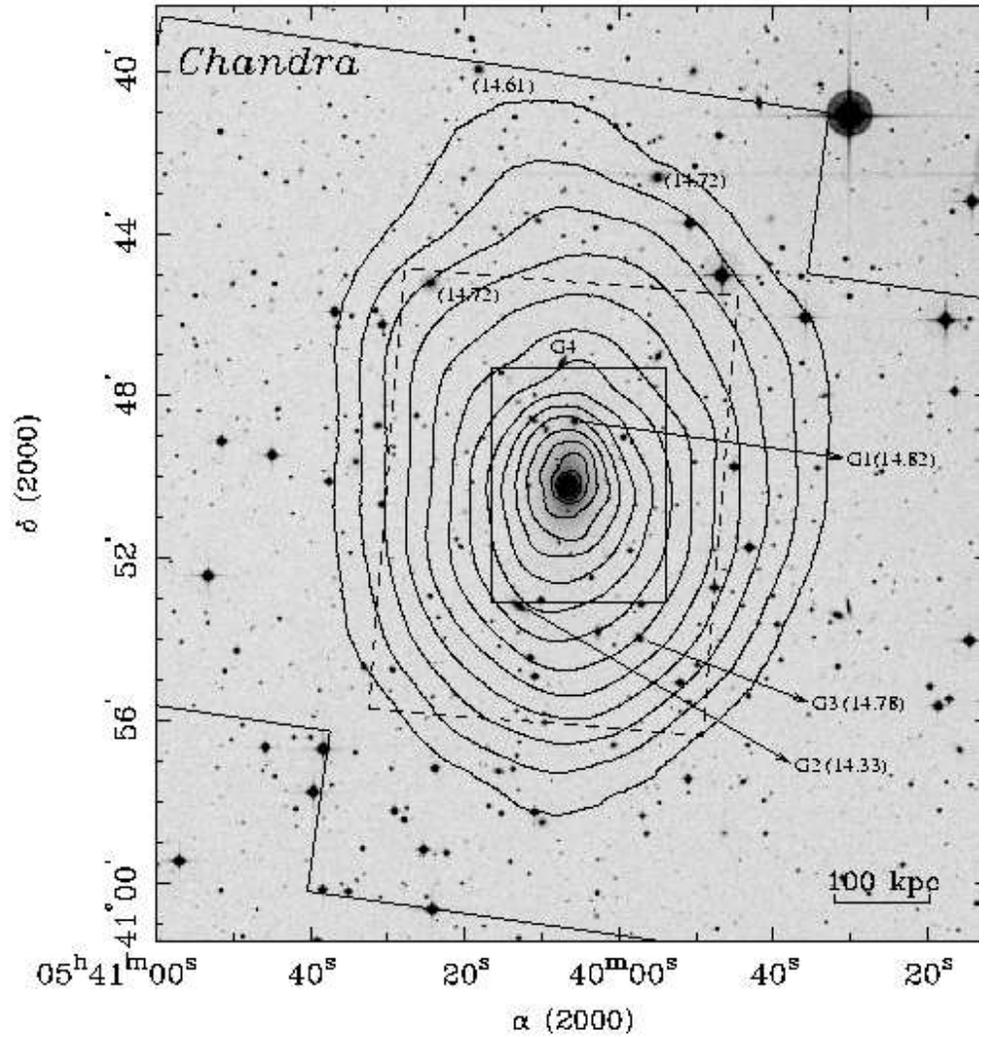}}
\vspace{0.2cm}
  \caption{\chandra\ 0.5 - 4 keV contours of the \esoa\ group emission (two pointings
combined) superposed on the DSS I image. The X-ray image was background-subtracted and
exposure-corrected. Point sources were replaced by averages of surrounding diffuse
emission. The X-ray image was then adaptively smoothed. Only the diffuse emission
in I chips are shown, while the emission on S2 chips is dominated by background.
The contours levels increase by a factor of $\sqrt{2}$ from the outermost one
(1.58$\times$10$^{-3}$ cts s$^{-1}$ arcmin$^{-2}$) towards the center. Group
emission beyond the outermost contour still exists but is very faint and affected
by the chip edges (shown as the solid lines). The box by solid
lines represents the zoom-in region in Fig. 7, while the box by dashed lines
represents the region shown in Fig. 10. All six galaxies with R $<$ 15 mag in
the field are marked by their magnitude. Galaxy G2, G3 and G4 are known group
members (Table 3).
    \label{fig:img:smo}}
\end{figure*}
\clearpage

\begin{figure*}
\vspace{-4.2cm}
  \centerline{\includegraphics[height=1.0\linewidth,angle=270]{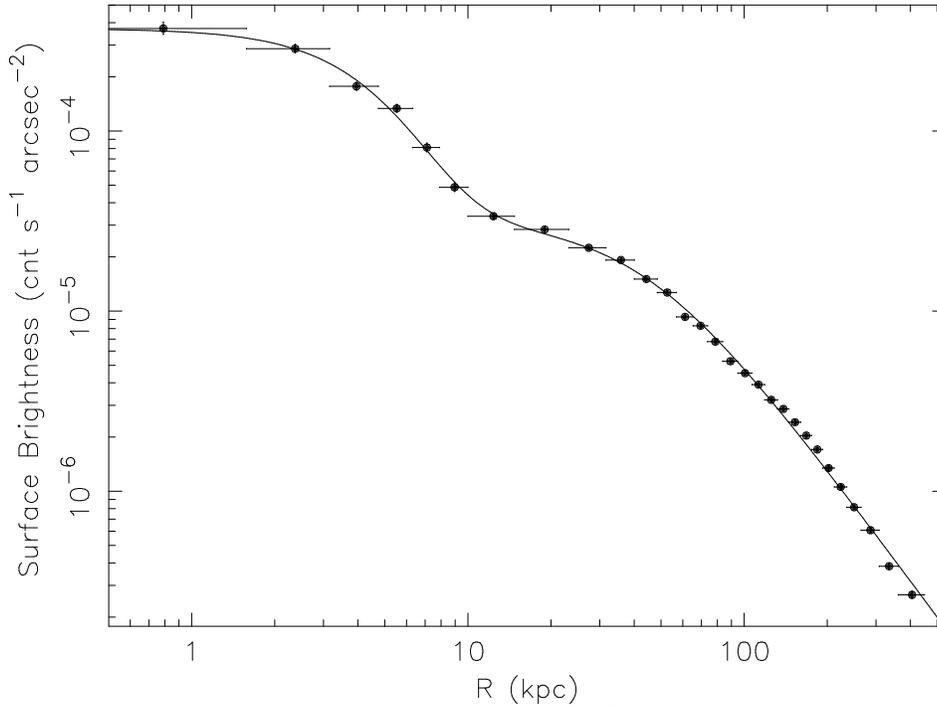}}
\vspace{-0.5cm}
  \caption{The \chandra\ 0.5 - 4 keV surface brightness profile of the \esoa\ group.
There are at least two components separated at 10 kpc. The best-fit double-$\beta$
model is also shown. Beyond 130 kpc, the surface brightness profile steepens
with an increase of $\sim$ 0.04 for $\beta$.
The surface brightness profile from \xmm, with worse resolution, is consistent
with that of \chandra\ in both normalization and shape.
   \label{fig:img:smo}}
\end{figure*}

\begin{figure*}
\vspace{-1cm}
  \centerline{\includegraphics[height=0.85\linewidth,angle=270]{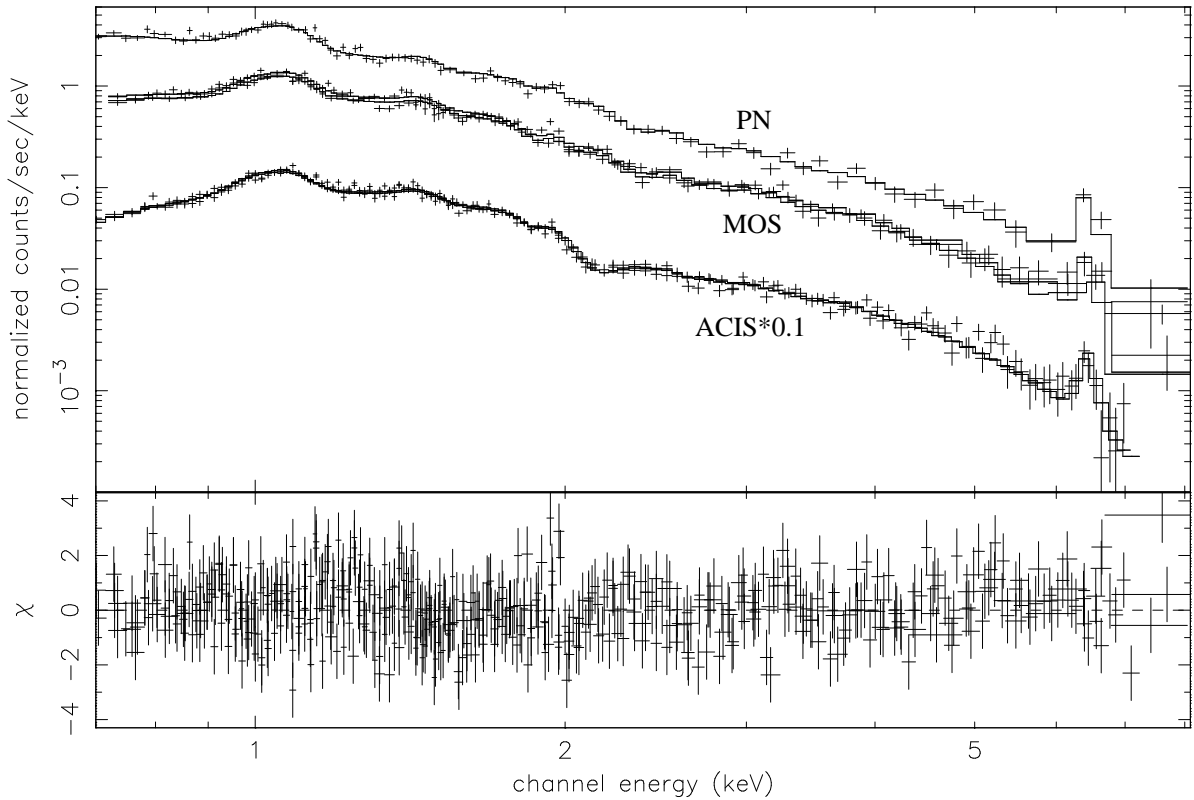}}
  \caption{The global spectra of \esoa\ within 0.33 \rv, including those of
two ACIS observations, EPIC MOS1, EPIC MOS2 and EPIC PN, are shown
with the best-fit VMEKAL model and residuals. The residuals are quite random.
Lines from Fe-L blend (1.0 - 1.3
keV), Si H-$\alpha$ ($\sim$ 2.0 keV) and Fe K$\alpha$ ($\sim$ 6.7 keV) are prominent.
    \label{fig:img:smo}}
\end{figure*}
\clearpage

\begin{figure*}
\vspace{-1.3cm}
  \centerline{\includegraphics[height=0.9\linewidth,angle=270]{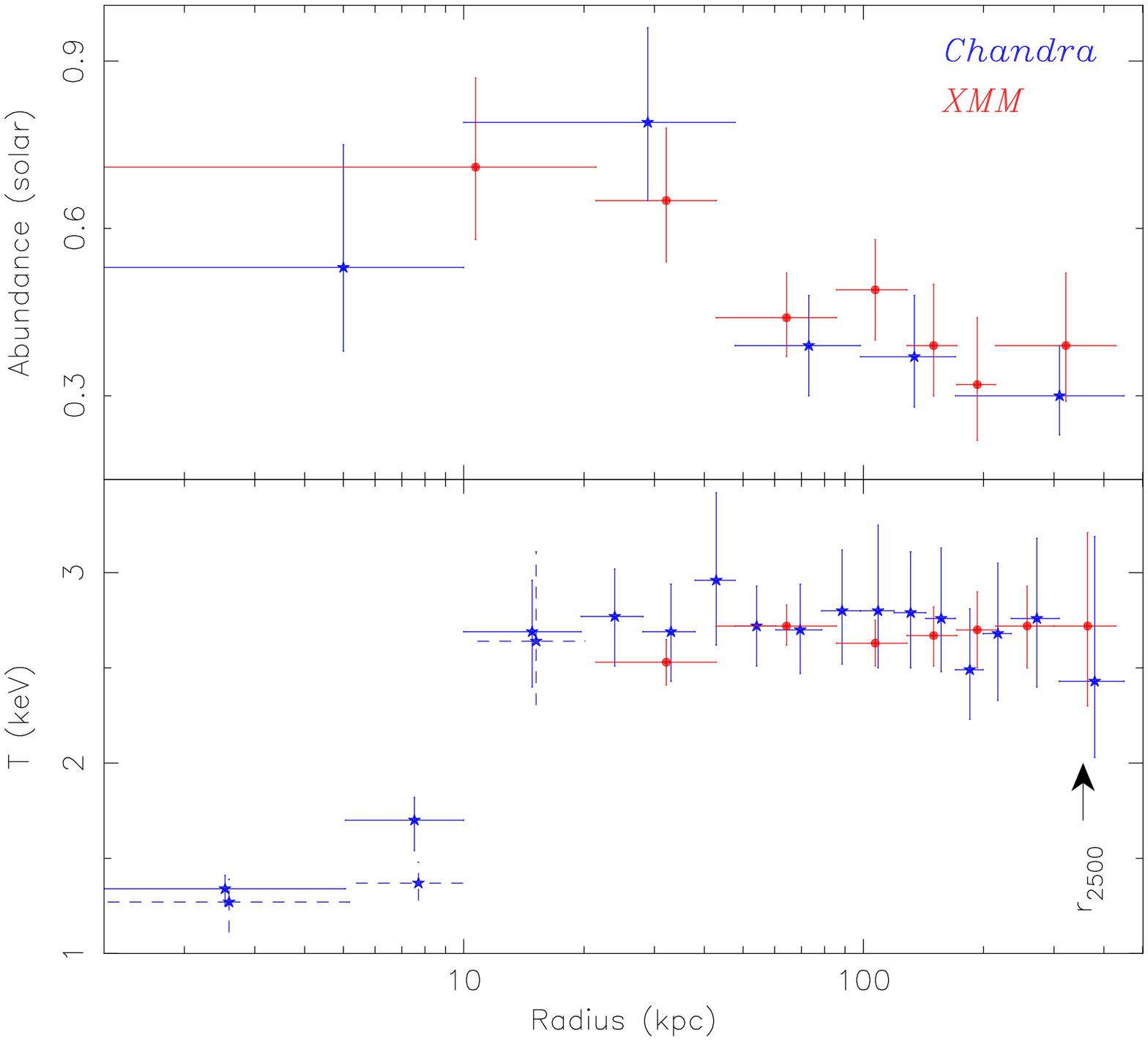}}
  \caption{Temperature and abundance profiles of the \esoa\ group, with
uncertainties represented by 90\% confidence levels. The red and
blue points represent the \chandra\ and the \xmm\ values respectively. The
PSF of \xmm\ and the highly peaked group emission affect the central part of
the \xmm\ temperature profile so that the central 30$''$ \xmm\ temperature is
not shown. However, the result is consistent with the averaged value of \chandra\
data. Apart from the PSF effect, the \chandra\ and \xmm\ results agree with
each other well. The dashed-line blue points of temperature within the central
20 kpc is the deprojected \chandra\ temperature at those regions.
The temperature profile is flat beyond the central 10 kpc to 450 kpc,
while the abundance is still not low ($\sim$ 0.35 solar) between 200 and 450 kpc
in this optically poor group.
    \label{fig:img:smo}}
\end{figure*}

\begin{figure*}
\vspace{-1.4cm}
  \centerline{\includegraphics[height=0.7\linewidth,angle=270]{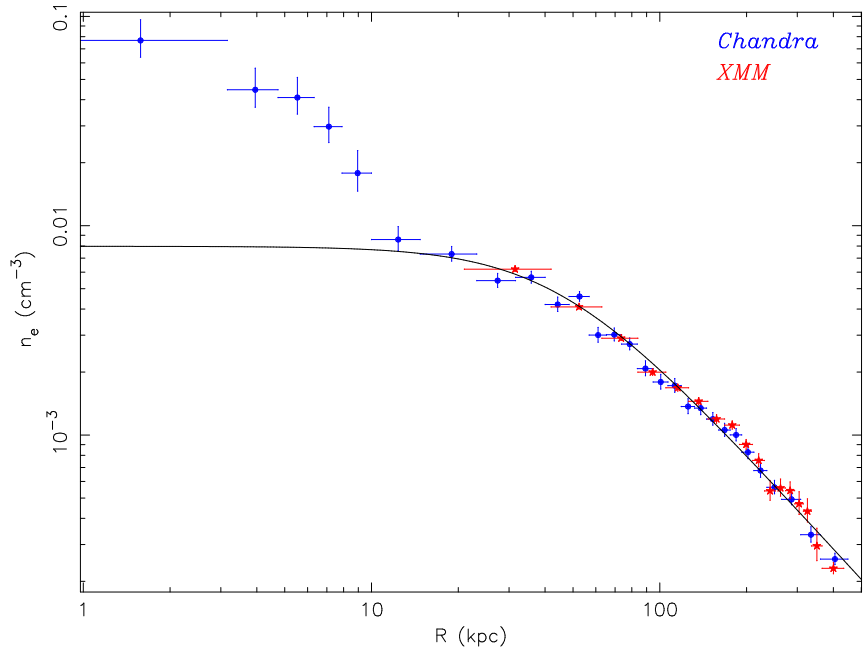}}
  \caption{\chandra\ and \xmm\ electron density profiles obtained from the
deprojection analysis (1 $\sigma$ uncertainty). They are consistent with each
other. The innermost \xmm\ bin is
not shown for clarity but is consistent with the average \chandra\ value.
The solid line is the best-fit $\beta$ model to all bins beyond the central
10 kpc. Aside from the central 10 kpc, the gas density distribution can be
described by a single $\beta$-model with $\beta$=0.504$\pm$0.011 and
r$_{0}$=44.4$\pm$2.6 kpc, but also shows steepening beyond $\sim$ 250 kpc.
    \label{fig:img:smo}}
\end{figure*}
\clearpage

\begin{figure*}
  \centerline{\includegraphics[height=1.0\linewidth,angle=270]{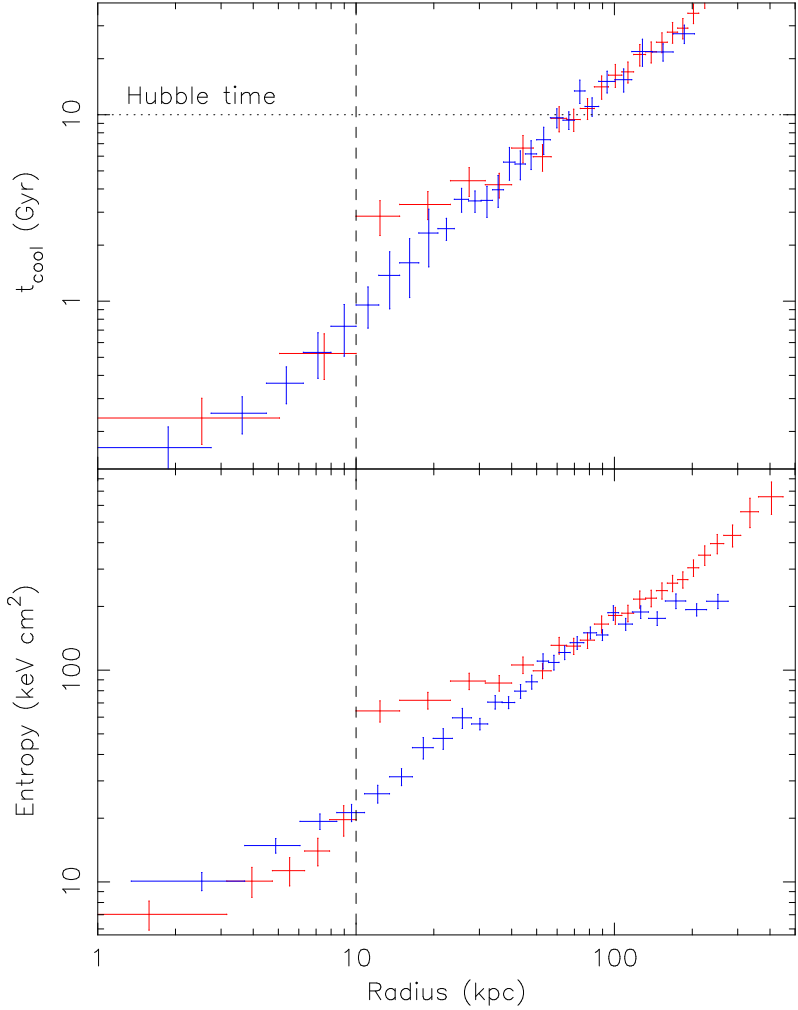}}
  \caption{Cooling time and entropy profiles of \esoa\ (in red), derived from electron
density profile and the best-fit to the temperature profile (see text).
The gas cooling time profile shows that everywhere within $\sim$ 50 kpc,
the gas cooling time is less than a Hubble time. We also plot the cooling time
profile of the NGC~1550 group (blue points; S03), which matched that of \esoa\
except the flattening region between 10 and 30 kpc. The scaled entropy profile
of the NGC~1550 group (scaled by the temperature, see text) (blue points; S03)
is also plotted, which is identical to that of \esoa\ except the excess of the
\esoa\ entropy between 10 and 50 kpc. This may imply that this entropy excess
region has been heated relative to that of the NGC~1550 group.
    \label{fig:img:smo}}
\end{figure*}
\clearpage

\begin{figure*}
\vspace{-8cm}
  \centerline{\includegraphics[height=1.3\linewidth]{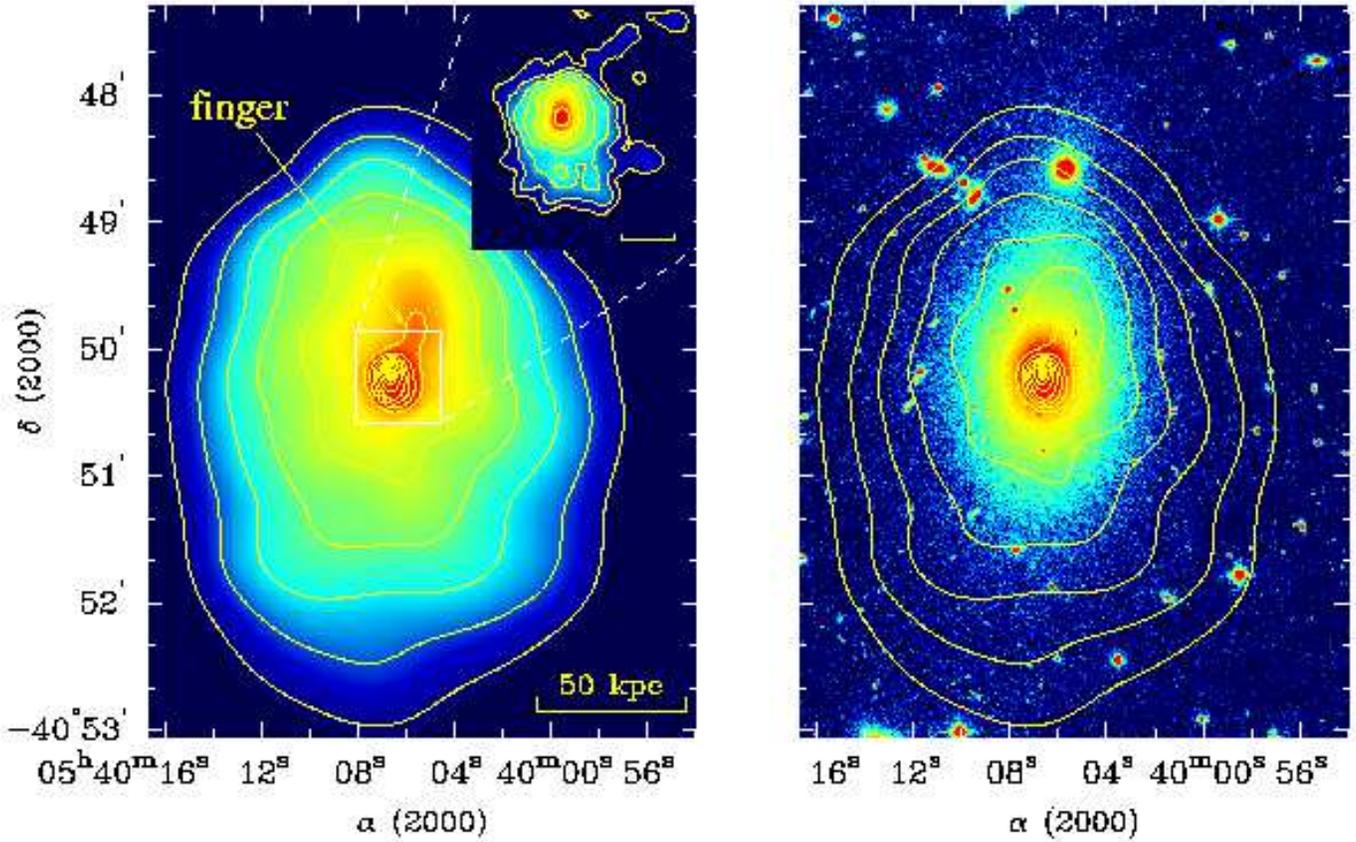}}
\vspace{1cm}
  \caption{{\bf Left}: ACIS-I 0.5 - 4 keV image (background-subtracted
and exposure-corrected) of the central 100 kpc region of the \esoa\ group. There
is a X-ray ``finger'' extending from the central dense component to the north.
The contours levels increase by a factor of $\sqrt{2}$ from the outermost
one (1.77$\times$10$^{-2}$ cts s$^{-1}$ arcmin$^{-2}$) towards the center.
The inner 15 kpc region is zoomed in the upper right. The surface brightness
decline to the north is especially sharp with a 10$''$ X-ray tail trailing
behind. The X-ray peak is also 2-3 kpc north of the isophotes center.
The contours levels increase by
a factor of $\sqrt{2}$ from the outermost one (1.36$\times$10$^{-2}$ cts
s$^{-1}$ arcsec$^{-2}$) towards the center. The scale bar represents 10 kpc.
{\bf Right}: the R-band image of the \esoa\ group (from the observations of
the Danish 1.54m telescope) in the same field superposed
with X-ray contours. The optical light is also elongated along the
north-south as the X-rays and has a faint extension to the north.
   \label{fig:img:smo}}
\end{figure*}
\clearpage

\begin{figure}
\vspace{-2cm}
  \centerline{\includegraphics[height=0.8\linewidth]{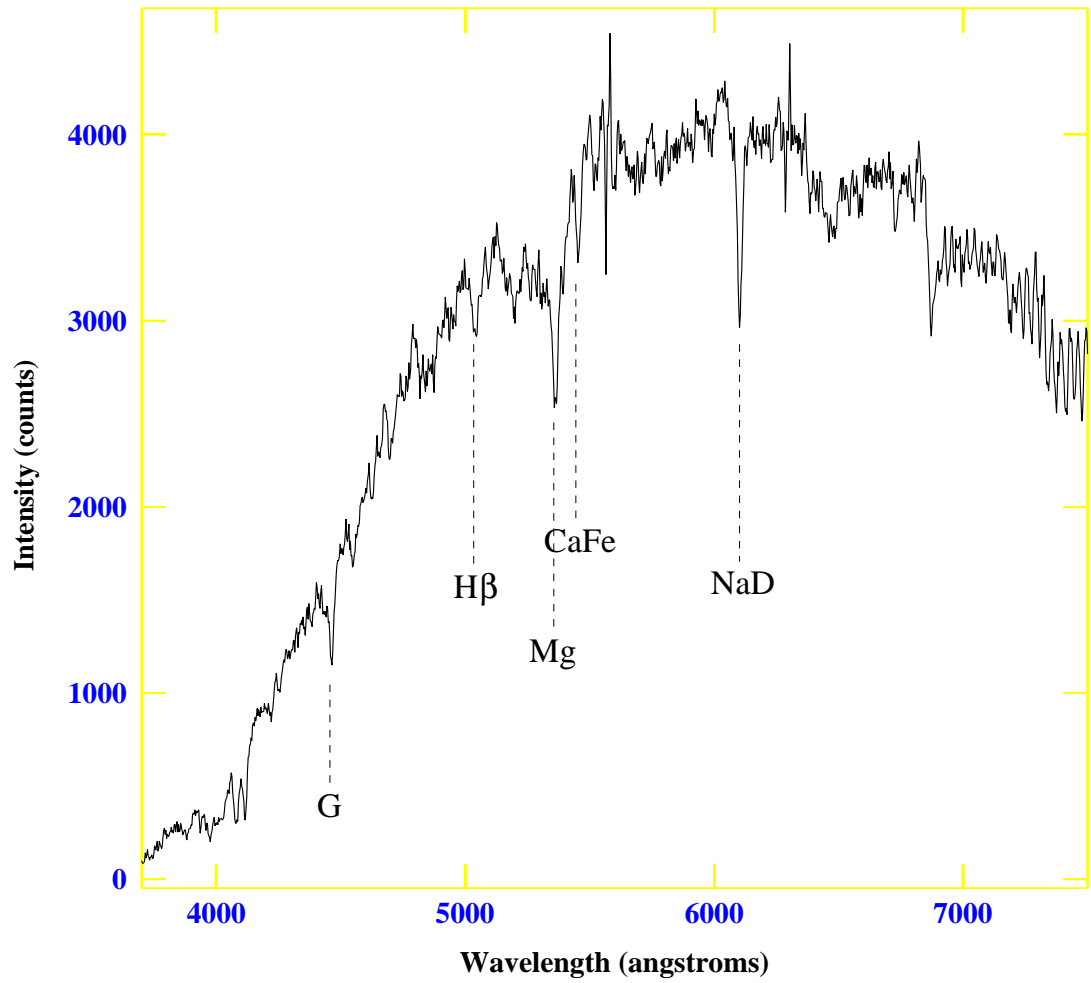}}
  \caption{The optical spectrum along the major axis of \esoa\ obtained with
the Danish 1.54 m telescope. Significant absorption lines are marked. No
significant emission lines are detected.
    \label{fig:img:smo}}
\end{figure}
\clearpage

\begin{figure*}
\vspace{-5cm}
  \centerline{\includegraphics[height=0.95\linewidth,angle=270]{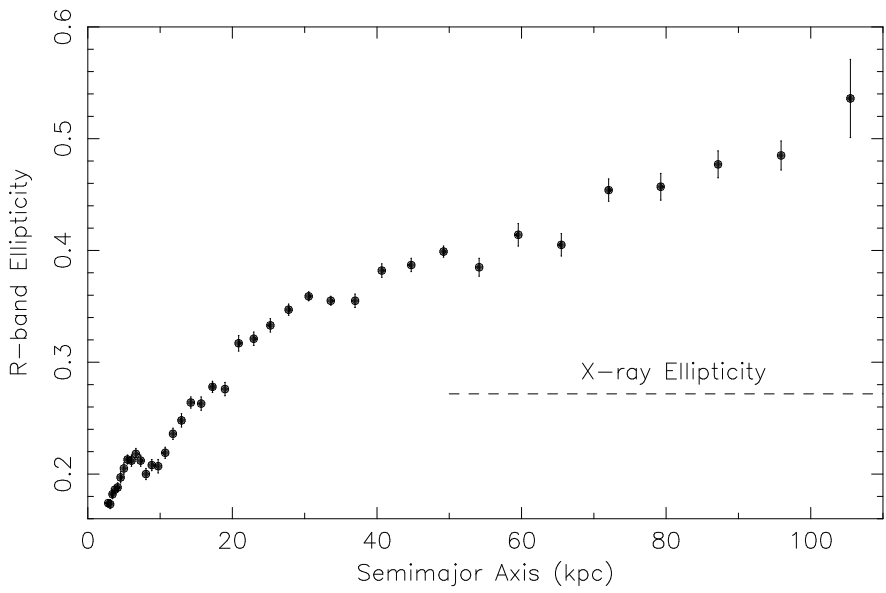}}
  \caption{Ellipticity of \esoa\ light measured in the R band, using ELLIPSE in
the STSDAS package. Ellipticity profile of the X-ray emission is flat with $\sim$
25\% random fluctuation. The average value is marked by the dashed line. The
optical ellipticity is significantly larger than that seen in the X-rays beyond
50 kpc.
    \label{fig:img:smo}}
\end{figure*}
\clearpage

\begin{figure*}
\vspace{-1cm}
  \centerline{\includegraphics[height=1.0\linewidth]{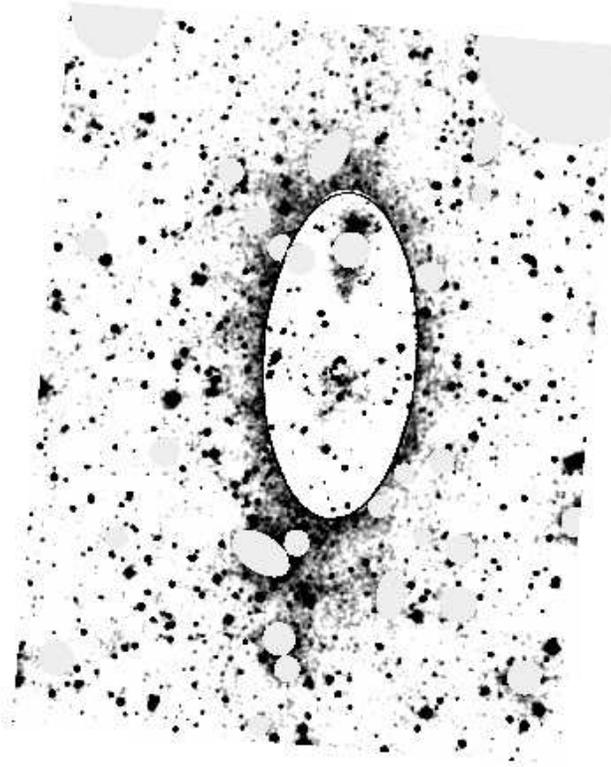}}
\vspace{0.5cm}
  \caption{The residual R-band light for \esoa\ after subtraction of a
two-dimensional elliptical model ($\S$4.2; Fig. 9). The big ellipse
(2.35$' \times$ 4.94$'$) represents the outermost region for which we
performed elliptical isophote fitting. A constant sky background was added to the
residuals inside the big ellipse for better presentation. Bright galaxies and
stars are masked. Within the big ellipse, residual
light from the best-fit model on the level of 3\% - 4\% is detected, around the
galaxy G1 and south of the galaxy center (to $\sim$ 50 kpc). The excess around
the galaxy G1 is likely to be produced by the tidal force of \esoa. At 130 -
200 kpc south of the center of \esoa\ (outside the big ellipse), significant
excess emission is also detected.
    \label{fig:img:smo}}
\end{figure*}
\clearpage

\begin{figure*}
\vspace{-1.2cm}
  \centerline{\includegraphics[height=0.8\linewidth,angle=270]{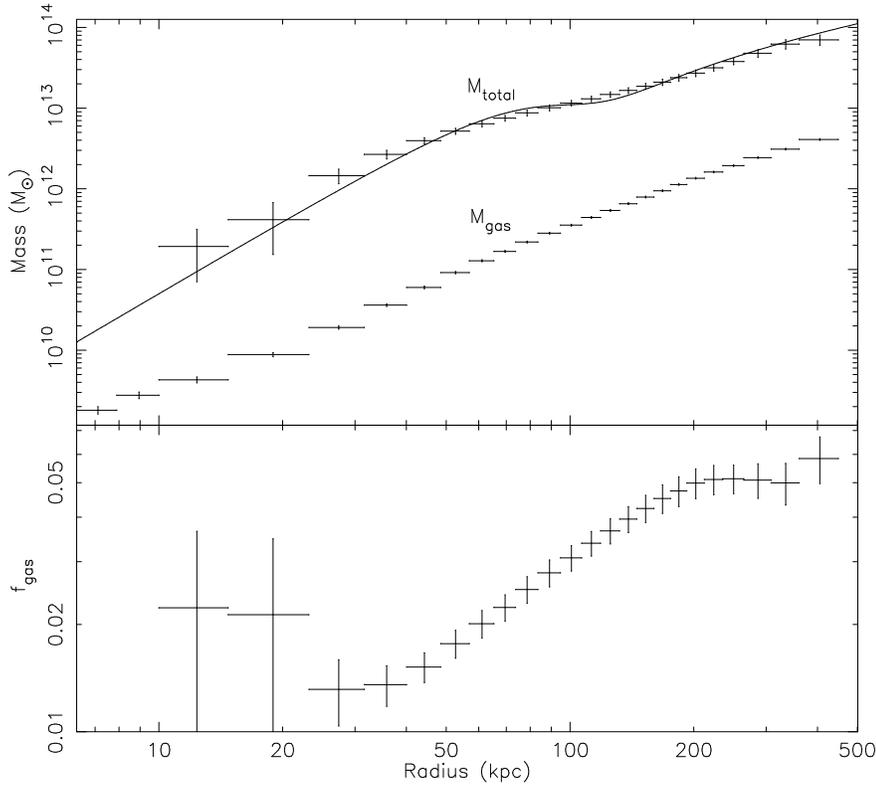}}
  \caption{{\bf Upper}: the total gravitational mass and gas mass profiles.
The solid line shows the total mass profile for isothermal gas with density
described by a double-$\beta$ model. {\bf Lower}: the gas fraction profile.
    \label{fig:img:smo}}
\end{figure*}

\begin{figure*}
\vspace{-1.6cm}
  \centerline{\includegraphics[height=0.75\linewidth,angle=270]{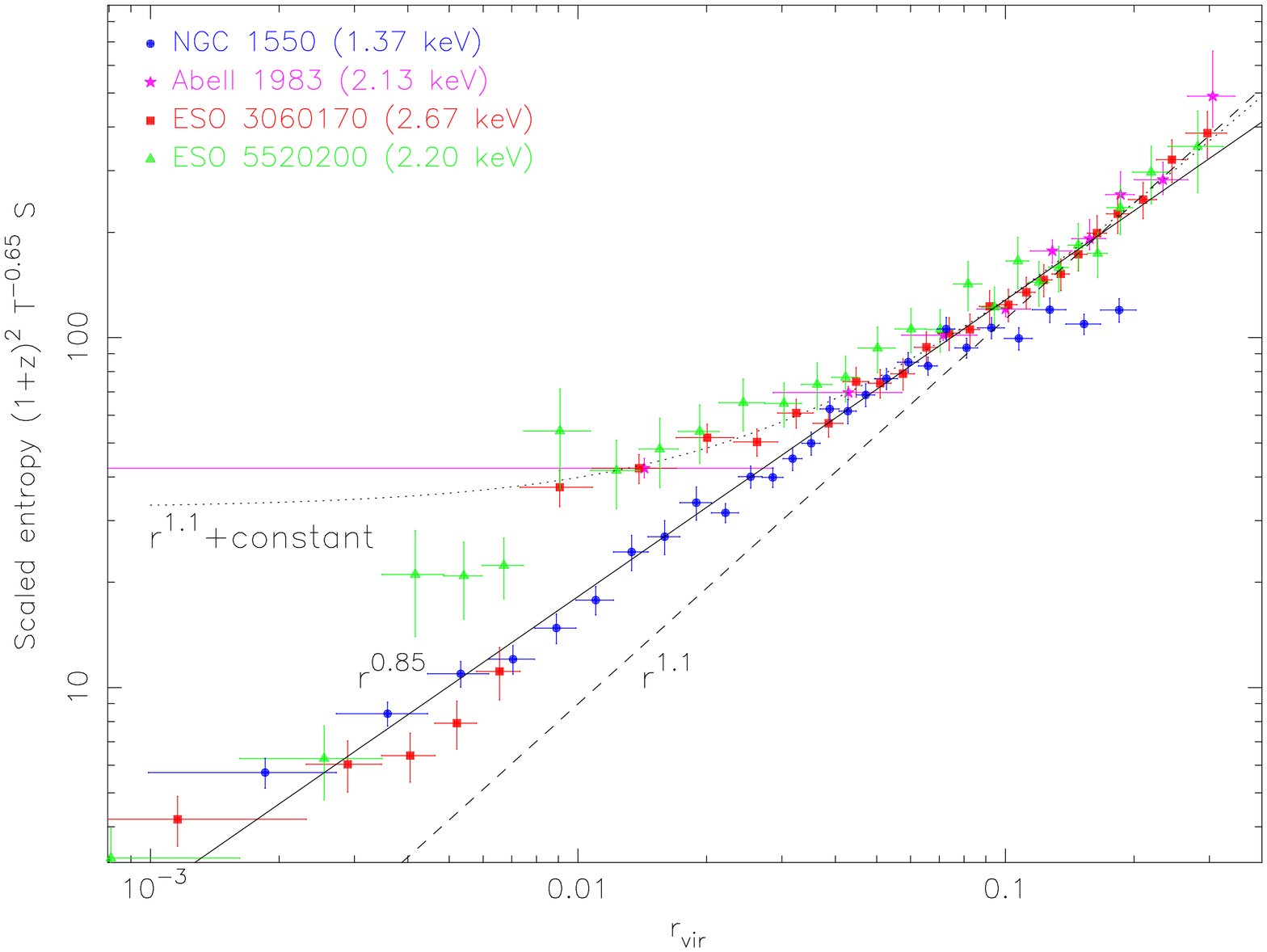}}
  \caption{Scaled entropy profiles of 3 hot nearby galaxy groups (T = 2-3 keV)
and a cool group with a normal cooling core (NGC~1550; S03).
The scaled entropy profiles of 2 - 3 keV groups (A1983, \esoa\ and \esob)
agree at large radii. Their entropy profiles flatten between 0.01 and 0.04
\rv. The profile of the NGC~1550 group, not only aligns with those of 2 -
3 keV hot groups in 0.04 - 0.09 \rv, but also extends to small radii with a
slope consistent with that of hot groups at large radii, while other three
hot groups show excess between 0.01 and 0.04 \rv\ compared to NGC~1550. Within
0.004 \rv, their scaled entropy values agree with each other again. The
variety of central entropy profiles of different groups may reflect the different
stage of their cooling cores. The solid line is the best fit (S
$\propto$ r$^{0.85}$) for the scaled entropy profiles excluding the flattened
ones between 0.01 and 0.04 \rv\ and the flattened profile of NGC~1550
beyond 0.09 \rv. The dashed line is the prediction from simulations involving
only gravity and shock heating (S $\propto$ r$^{1.1}$ by Tozzi \& Norman
2001) and it is normalized to match the observed entropy between 0.2 and 0.3 \rv.
The flattening of the entropy profiles within 0.1 \rv should be related to the
non-gravitational processes. The dotted line is the best fit of the model:
r$^{1.1}$ + constant (representing an isentropic core) for the entropy of the
hot groups beyond 0.007 \rv. It well fits the data beyond the central cool core.
    \label{fig:img:smo}}
\end{figure*}

\end{document}